
\magnification=1200
\voffset=0 true mm
\hoffset=0 true in
\hsize=6.5 true in
\vsize=8.5 true in
\normalbaselineskip=13pt
\def\doublespace{\baselineskip=20pt plus 3pt\message{double space}}
\def\singlespace{\baselineskip=13pt\message{single space}}
\let\spacing=\singlespace
\parindent=1.0 true cm



\newcount\equationumber \newcount\sectionumber
\sectionumber=1 \equationumber=1
\def\setsection{\global\advance\sectionumber by1 \equationumber=1}

\def\numbe{{{\number\sectionumber}{.}\number\equationumber}
                            \global\advance\equationumber by1}
\def\numberit{\eqno{(\number\equationumber)} \global\advance\equationumber by1}

\def\numberal{(\number\equationumber)\global\advance\equationumber by1}

\def\sectionit{\eqno{(\numbe)}}

\def\ccf#1{\,\vcenter{\normalbaselines
    \ialign{\hfil$##$\hfil&&$\>\hfil ##$\hfil\crcr
      \mathstrut\crcr\noalign{\kern-\baselineskip}
      #1\crcr\mathstrut\crcr\noalign{\kern-\baselineskip}}}\,}
\def\scf#1{\,\vcenter{\baselineskip=9pt
    \ialign{\hfil$##$\hfil&&$\>\hfil ##$\hfil\crcr
      \vphantom(\crcr\noalign{\kern-\baselineskip}
      #1\crcr\mathstrut\crcr\noalign{\kern-\baselineskip}}}\,}

\def\small3j#1#2#3#4#5#6{\def\st{\scriptstyle} 
   \bigl(\scf{\st#1&\st#2&\st#3\cr
           \st#4&\st#5&\st#6\cr} \bigr)}




\def\ref#1{$^{#1)}$}


\def\upa#1{\raise 1pt\hbox{\sevenrm #1}}
\def\dna#1{\lower 1pt\hbox{\sevenrm #1}}
\def\dnb#1{\lower 2pt\hbox{$\scriptstyle #1$}}
\def\dnc#1{\lower 3pt\hbox{$\scriptstyle #1$}}
\def\upb#1{\raise 2pt\hbox{$\scriptstyle #1$}}
\def\upc#1{\raise 3pt\hbox{$\scriptstyle #1$}}
\def\hprime{\raise 2pt\hbox{$\scriptstyle \prime$}}
\def\ccom{\,\raise2pt\hbox{,}}


\def\asymptotically#1{\;\rlap{\lower 4pt\hbox to 2.0 true cm{
    \hfil\sevenrm  #1 \hfil}}
   \hbox{$\relbar\joinrel\relbar\joinrel\relbar\joinrel
     \relbar\joinrel\relbar\joinrel\longrightarrow\;$}}
\def\Asymptotically#1{\;\rlap{\lower 4pt\hbox to 3.0 true cm{
    \hfil\sevenrm  #1 \hfil}}
   \hbox{$\relbar\joinrel\relbar\joinrel\relbar\joinrel\relbar\joinrel
     \relbar\joinrel\relbar\joinrel\relbar\joinrel\relbar\joinrel
     \relbar\joinrel\relbar\joinrel\longrightarrow$\;}}

\def\dal{\mathop{\sqcup\hskip-6.4pt\sqcap}\nolimits}

\catcode`@=11
\def\C@ncel#1#2{\ooalign{$\hfil#1\mkern2mu/\hfil$\crcr$#1#2$}}
\def\gf#1{\mathrel{\mathpalette\c@ncel#1}}      
\def\Gf#1{\mathrel{\mathpalette\C@ncel#1}}      

\def\gapx{\lower 2pt \hbox{$\buildrel>\over{\scriptstyle{\sim}}$}}
\def\lapx{\lower 2pt \hbox{$\buildrel<\over{\scriptstyle{\sim}}$}}

\def\nablaleft{\hbox{\raise 6pt\rlap{{\kern-1pt$\leftarrow$}}{$\nabla$}}}
\def\nablaright{\hbox{\raise 6pt\rlap{{\kern-1pt$\rightarrow$}}{$\nabla$}}}
\def\nablaboth{\hbox{\raise 6pt\rlap{{\kern-1pt$\leftrightarrow$}}{$\nabla$}}}

\def\boks#1#2{{\hsize=#1 true cm\parindent=0pt
  {\vbox{\hrule height1pt \hbox
    {\vrule width1pt \kern3pt\raise 3pt\vbox{\kern3pt{#2}}\kern3pt
    \vrule width1pt}\hrule height1pt}}}}

\def\heading{ }
\def\range{ }

\def\body{\vfill\eject\parindent=1.0 true cm
 \ifx\spacing\singlespace\singlespace\else\doublespace\fi}
\def\title#1{\centerline{{\bf #1}}}

\def\today{\ifcase\month\or
  January\or February\or March\or April\or May\or June\or
  July\or August\or September\or October\or November\or December\fi
  \space\number\day, \number\year}
\let\date=\today
\newcount\hour \newcount\minute
\countdef\hour=56
\countdef\minute=57
\hour=\time
  \divide\hour by 60
  \minute=\time
  \count58=\hour
  \multiply\count58 by 60
  \advance\minute by -\count58

\def\sectionskip{\penalty-500\vskip24pt plus12pt minus6pt}

\def\sec{\hbox{\lower 1pt\rlap{{\sixrm S}}{\hbox{\raise 1pt\hbox{\sixrm S}}}}}
\def\section#1\par{\goodbreak\message{#1}
    \sectionskip\nobreak\noindent{\bf #1}\vskip0.3cm \noindent}

\nopagenumbers
\headline={\ifnum\pageno=\count31\frontheadline
  \else{\ifnum\pageno=0\frontheadline
     \else{{\raise 2pt\hbox to \hsize{\paperhead}}}\fi}\fi}

\footline={\centerline{\sevenbf \folio}}
\def\frontheadline{\sevenbf \hfil}
\def\paperhead{\sevenbf \heading\ \range\hfil\folio}
\newdimen\pagewidth \newdimen\pageheight \newdimen\ruleht
\maxdepth=2.2pt
\pagewidth=\hsize \pageheight=\vsize \ruleht=.5pt

\def\onepageout#1{\shipout\vbox{ 
    \offinterlineskip 
  \makeheadline
    \vbox to \pageheight{
         #1 
 \ifnum\pageno=\count31{\vskip 21pt\line{\the\footline}}\fi
 \ifvoid\footins\else 
 \vskip\skip\footins \kern-3pt
 \hrule height\ruleht width\pagewidth \kern-\ruleht \kern3pt
 \unvbox\footins\fi
 \boxmaxdepth=\maxdepth}
 \advancepageno}}
\output{\onepageout{\pagecontents}}
\count31=-1
\topskip 0.7 true cm

\doublespace
\pageno=0
\centerline{\bf Consistency of the Nonsymmetric Gravitational Theory}
\centerline{\bf J. W. Moffat}
\centerline{\bf Department of Physics}
\centerline{\bf University of Toronto}
\centerline{\bf Toronto, Ontario M5S 1A7}
\centerline{\bf Canada}
\vskip 2 true in
\centerline{\bf October 1994}
\vskip 2.5 true in
{\bf UTPT-94-30. gr-qc/9411027}
\vskip 0.3 true in
e-mail: moffat@medb.physics.utoronto.ca
\par\vfil\eject
\centerline{\bf Consistency of the Nonsymmetric Gravitational Theory}
\centerline{\bf J. W. Moffat}
\centerline{\bf Department of Physics}
\centerline{\bf University of Toronto}
\centerline{\bf Toronto, Ontario M5S 1A7}
\centerline{\bf Canada}
\vskip 0.3 true in
\centerline{\bf Abstract}
\vskip 0.2 true in
A nonsymmetric gravitational theory (NGT) is presented
which is free of ghost
poles, tachyons and higher-order poles and there are no problems with
asymptotic boundary conditions. An extended Birkhoff theorem is shown to
hold for the spherically symmetric solution of the field equations.
A static spherically symmetric solution in the short-range approximation,
$\mu^{-1} > 2m$, is everywhere regular and does not contain a black hole event
horizon.
\par\vfil\eject
\proclaim 1. {\bf Introduction}\par
\vskip 0.2 true in
An earlier version of the nonsymmetric gravitational theory (NGT)$^{1}$,
based on a nonsymmetric field structure with the decompositions$^{2}$:
$$
g_{(\mu\nu)}={1\over 2}(g_{\mu\nu}+g_{\nu\mu}),\quad g_{[\mu\nu]}=
{1\over 2}(g_{\mu\nu}-g_{\nu\mu}),
\sectionit
$$
and
$$
\Gamma^\lambda_{\mu\nu}=\Gamma^\lambda_{(\mu\nu)}
+\Gamma^\lambda_{[\mu\nu]},
\sectionit
$$
was confronted with consistency problems due to the absence of a massless gauge
invariance in the antisymmetric sector of the theory. The only gauge
invariance was associated with the general covariance of the theory.
Attempts to derive a perturbative NGT scenario free of couplings to unphysical
modes and with consistent asymptotic boundary conditions proved to be a
difficult (if not impossible) task$^{3,4}$.

In the following, a version of NGT is presented which will be shown
to be free of ghosts, tachyons and higher-order poles in the propagator in the
linear approximation$^{5}$. An expansion of $g_{\mu\nu}$ about an arbitrary
Einstein
background metric also yields field equations to first order in $g_{[\mu\nu]}$,
which are free of couplings to unphysical (negative energy) modes; the
solutions
of the field equations have consistent asymptotic boundary conditions. This is
accomplished by adding two cosmological terms to the earlier version of the NGT
Lagrangian density.

In view of the difficulty in obtaining physically consistent geometrical
generalizations of Einstein gravitational theory (EGT), it is interesting that
such a consistent theory can
be formulated. For example, higher derivative (higher powers of the
scalar curvature $R$) theories possess ghost poles, tachyons and higher-order
poles in the linear approximation$^{6}$.

The static spherically symmetric solution of the NGT field equations
is shown to satisfy a Birkhoff theorem.
A static spherically symmetric solution in the short-range approximation,
$\mu^{-1} > 2m$, is everywhere regular and does not contain a black hole
event horizon$^{7}$.
\vskip 0.2 true in
\setsection\proclaim 2. {\bf Nonsymmetric Gravitational Theory}\par
\vskip 0.2 true in
The non-Riemannian geometry is based on the nonsymmetric field structure with a
nonsymmetric $g_{\mu\nu}$ and affine connection $\Gamma^\lambda_{\mu\nu}$,
defined in Eqs.(1.1) and (1.2).
The contravariant tensor $g^{\mu\nu}$ is defined in terms of the equation:
$$
g^{\mu\nu}g_{\sigma\nu}=g^{\nu\mu}g_{\nu\sigma}=\delta^\mu_\sigma.
\sectionit
$$

The Lagrangian density is given by
$$
{\cal L}_{NGT}={\cal L}_R+{\cal L}_M,
\sectionit
$$
where
$$
{\cal L}_R={\bf g}^{\mu\nu}R_{\mu\nu}(W)-2\lambda\sqrt{-g}
-{1\over 4}\mu^2{\bf g}^{\mu\nu}g_{[\nu\mu]}+{1\over 2}\sigma {\bf g}^{\mu\nu}
W_\mu W_\nu,
\sectionit
$$
where $\lambda$ is the cosmological constant and $\mu^2$ and $\sigma$ are
additional cosmological constants associated with $g_{[\mu\nu]}$ and $W_\mu$,
respectively. Moreover, ${\cal L}_M$ is the matter Lagrangian density
($G=c=1$):
$$
{\cal L}_M=-8\pi g^{\mu\nu}{\bf T}_{\mu\nu}.
\sectionit
$$
Here, ${\bf g}^{\mu\nu}=\sqrt{-g}g^{\mu\nu}$ and $R_{\mu\nu}(W)$ is the
NGT contracted curvature tensor:
$$
R_{\mu\nu}(W)=W^\beta_{\mu\nu,\beta} - {1\over
2}(W^\beta_{\mu\beta,\nu}+W^\beta_{\nu\beta,\mu}) -
W^\beta_{\alpha\nu}W^\alpha_{\mu\beta} +
W^\beta_{\alpha\beta}W^\alpha_{\mu\nu},
\sectionit
$$
defined in terms of the unconstrained nonsymmetric connection:
$$
W^\lambda_{\mu\nu}=\Gamma^\lambda_{\mu\nu}-{2\over 3}\delta^\lambda_\mu
W_\nu,
\sectionit
$$
where
$$
W_\mu={1\over 2}(W^\lambda_{\mu\lambda}-W^\lambda_{\lambda\mu}).
\sectionit
$$
Eq.(2.6) leads to the result:
$$
\Gamma_\mu=\Gamma^\lambda_{[\mu\lambda]}=0.
\sectionit
$$
Moreover, ${\bf T}_{\mu\nu}$ is the matter source density tensor.

The NGT contracted curvature tensor can be written as
$$
R_{\mu\nu}(W) = R_{\mu\nu}(\Gamma) + {2\over 3} W_{[\mu,\nu]},
\sectionit
$$
where $R_{\mu\nu}(\Gamma)$ is defined by
$$
R_{\mu\nu}(\Gamma ) = \Gamma^\beta_{\mu\nu,\beta} -{1\over 2}
\left(\Gamma^\beta_{(\mu\beta),\nu} + \Gamma^\beta_{(\nu\beta),\mu}\right) -
\Gamma^\beta_{\alpha\nu} \Gamma^\alpha_{\mu\beta} +
\Gamma^\beta_{(\alpha\beta)}\Gamma^\alpha_{\mu\nu}.
\sectionit
$$

The field equations in the presence of matter sources are given by:
$$
G_{\mu\nu} (W)+\lambda g_{\mu\nu}+{1\over 4}\mu^2C_{\mu\nu}+{1\over 2}
\sigma(P_{\mu\nu}-{1\over 2}g_{\mu\nu}P) = 8\pi T_{\mu\nu},
\sectionit
$$
$$
{{\bf g}^{[\mu\nu]}}_{,\nu} = 3{\bf D}^\mu,
\sectionit
$$
$$
{{\bf g}^{\mu\nu}}_{,\sigma}+{\bf g}^{\rho\nu}W^\mu_{\rho\sigma}
+{\bf g}^{\mu\rho}
W^\nu_{\sigma\rho}-{\bf g}^{\mu\nu}W^\rho_{\sigma\rho}
+{2\over 3}\delta^\nu_\sigma{\bf g}^{\mu\rho}W^\beta_{[\rho\beta]}
$$
$$
+{\bf D}^\nu\delta^\mu_\sigma-{\bf D}^\mu\delta^\nu_\sigma=0.
\sectionit
$$
Here, we have
$$
G_{\mu\nu} = R_{\mu\nu} - {1\over 2} g_{\mu\nu} R,
\sectionit
$$
$$
C_{\mu\nu}=g_{[\mu\nu]}+{1\over 2}g_{\mu\nu}g^{[\sigma\rho]}
g_{[\rho\sigma]}+g^{[\sigma\rho]}g_{\mu\sigma}g_{\rho\nu},
\sectionit
$$
$$
P_{\mu\nu}=W_\mu W_\nu,
\sectionit
$$
and $P=g^{\mu\nu}P_{\mu\nu}=g^{(\mu\nu)}W_\mu W_\nu$. Also, we have
$$
{\bf D}^\mu={1\over 2}\sigma{\bf g}^{(\mu\alpha)}W_\alpha,
\sectionit
$$
and from (2.12), it follows that ${{\bf D}^\mu}_{,\mu}=0$.

We can write the field equations (2.11) in the form:
$$
R_{\mu\nu}(W)=\lambda g_{\mu\nu}+8\pi {\tilde T}_{\mu\nu},
\sectionit
$$
where
$$
{\tilde T}_{\mu\nu}=T_{\mu\nu}-{1\over 2}g_{\mu\nu}T
-{1\over 32\pi}\mu^2I_{\mu\nu}-{1\over 16\pi}\sigma P_{\mu\nu},
\sectionit
$$
and $T=g^{\mu\nu}T_{\mu\nu}$. Moreover, we have
$$
I_{\mu\nu}=C_{\mu\nu}-{1\over 2}g_{\mu\nu}C
=g^{[\sigma\rho]}g_{\mu\sigma}g_{\rho\nu}+{1\over 2}g_{\mu\nu}
g_{[\sigma\rho]}g^{[\sigma\rho]}+g_{[\mu\nu]}.
\sectionit
$$

In empty space, the field equations (2.18) become:
$$
R_{\mu\nu}(\Gamma)={2\over 3}W_{[\nu,\mu]}+\lambda
g_{\mu\nu}-{1\over 4}\mu^2I_{\mu\nu}-{1\over 2}\sigma P_{\mu\nu}.
\sectionit
$$

Substituting (2.6) into (2.13), we get
$$
{{\bf g}^{\mu\nu}}_{,\sigma}+{\bf g}^{\rho\nu}
\Gamma^\mu_{\rho\sigma}+{\bf g}^{\mu\rho}\Gamma^\nu_{\sigma\rho}
-{\bf g}^{\mu\nu}\Gamma^\rho_{(\sigma\rho)}
+{\bf D}^\nu\delta^\mu_\sigma-{\bf D}^\mu\delta^\nu_\sigma=0.
\sectionit
$$
This equation may be written in the form:
$$
g_{\mu\nu,\sigma}-g_{\rho\nu}\Gamma^\rho_{\mu\sigma}-g_{\mu\rho}
\Gamma^\rho_{\sigma\nu}=D^\rho\Lambda_{\mu\nu\sigma\rho},
\sectionit
$$
where
$$
\Lambda_{\mu\nu\sigma\rho}=g_{\mu\rho}g_{\sigma\nu}-g_{\mu\sigma}g_{\rho\nu}
-g_{\mu\nu}g_{[\sigma\rho]},
\sectionit
$$
and we have:
$$
{\sqrt{-g}}_{,\sigma}-\sqrt{-g}\Gamma^\rho_{(\sigma\rho)}
=\sqrt{-g}g_{[\rho\sigma]}D^\rho.
\sectionit
$$

The generalized Bianchi identities:
$$
[{\bf g}^{\alpha\nu}G_{\rho\nu}(\Gamma)+{\bf g}^{\nu\alpha}
G_{\nu\rho}(\Gamma)]_{,\alpha}+{g^{\mu\nu}}_{,\rho}{\bf G}_{\mu\nu}=0,
\sectionit
$$
give rise to the matter response equations:
$$
g_{\mu\rho}{{\bf T}^{\mu\nu}}_{,\nu}+g_{\rho\mu}{{\bf T}^{\nu\mu}}_{,\nu}
+(g_{\mu\rho,\nu}+g_{\rho\nu,\mu}-g_{\mu\nu,\rho}){\bf T}^{\mu\nu}=0.
\sectionit
$$
\vskip 0.2 true in
\setsection\proclaim 3. {\bf Linear Approximation}\par
\vskip 0.2 true in
Let us assume that $\lambda=0$ and expand $g_{\mu\nu}$ about Minkowski
spacetime:
$$
g_{\mu\nu}=\eta_{\mu\nu}+{}^{(1)}h_{\mu\nu}+...,
\sectionit
$$
where $\eta_{\mu\nu}$ is the Minkowski metric tensor: $\eta_{\mu\nu}=
\hbox{diag}(-1, -1, -1, +1)$. We also expand $\Gamma^\lambda_{\mu\nu}$ and
$W^\lambda_{\mu\nu}$:
$$
\Gamma^\lambda_{\mu\nu}={}^{(1)}\Gamma^\lambda_{\mu\nu}
+{}^{(2)}\Gamma^\lambda_{\mu\nu}+...,
$$
$$
W_\mu={}^{(1)}W_\mu+{}^{(2)}W_\mu+...\,.
\sectionit
$$
 We find from (2.1) that
$$
g^{\mu\nu}=\eta^{\mu\nu}-{}^{(1)}h^{\mu\nu}+...,
\sectionit
$$
where $h^{\mu\nu}=\eta^{\mu\lambda}\eta^{\sigma\nu}h_{\sigma\lambda}$.
Let us adopt the notation: $\psi_{\mu\nu}={}^{(1)}h_{[\mu\nu]}$. To first order
of
approximation, Eq.(2.12) gives
$$
\psi_\mu={3\over 2}\sigma W_\mu,
\sectionit
$$
where for convenience $W_\mu$ denotes ${}^{(1)}W_\mu$. Moreover,
$$
\psi_\mu={\psi_{\mu\beta}}^{,\beta}=\eta^{\beta\sigma}\psi_{\mu\beta,\sigma}.
\sectionit
$$
{}From Eqs.(2.23), we obtain the result to first order:
$$
{}^{(1)}{\Gamma^\lambda_{\mu\nu}}={1\over 2}\eta^{\lambda\sigma}
({}^{(1)}h_{\sigma\nu,\mu}
+{}^{(1)}h_{\mu\sigma,\nu}-{}^{(1)}h_{\nu\mu,\sigma})+{1\over 2}\sigma
({\delta^\lambda_\mu} W_\nu-{\delta^\lambda_\nu} W_\mu).
\sectionit
$$

The antisymmetric and symmetric field equations derived from Eq.(2.11)
decouple to lowest order; the symmetric equations are the usual
Einstein field equations in the linear approximation. The skew
equations are given by
$$
(\dal+\mu^2)\psi_{\mu\nu}=\kappa W_{[\nu,\mu]}+16\pi T_{[\mu\nu]},
\sectionit
$$
where $\kappa=\sigma+{4\over 3}$. For $\sigma=\mu=0$, Eq.(3.7) reduces to the
linearized antisymmetric equations of the earlier version of NGT$^{1}$:
$$
\dal\psi_{\mu\nu}={4\over 3}W_{[\nu,\mu]}+16\pi T_{[\mu\nu]}.
\sectionit
$$

If we take the divergence of Eq.(3.7), we get
$$
\tau\dal W_\mu+{3\over 2}\sigma\mu^2W_\mu=16\pi {T_{[\mu\nu]}}^{,\nu},
\sectionit
$$
where $\tau=2(\sigma+{1\over 3})$. Let us choose $\tau=0$ corresponding to
$\sigma=-1/3$ which gives
$$
W_\mu=-{32\pi\over \mu^2}{T_{[\mu\nu]}}^{,\nu}.
\sectionit
$$
Substituting this into (3.8) yields the Proca inhomogeneous
Klein-Gordon equation:
$$
(\dal+\mu^2)\psi_{\mu\nu}=J_{\mu\nu},
\sectionit
$$
where
$$
J_{\mu\nu}=16\pi(T_{[\mu\nu]}+{2\over \mu^2}{T_{[[\mu\sigma],\nu]}}^{,\sigma}).
\sectionit
$$
Moreover, from Eq.(3.4) we get
$$
\psi_\mu={16\pi\over \mu^2}{T_{[\mu\nu]}}^{,\nu}.
\sectionit
$$

Taking the divergence of (3.12) gives
$$
{J_{\mu\nu}}^{,\nu}=16\pi({T_{[\mu\nu]}}^{,\nu}+{1\over \mu^2}\dal
{T_{[\mu\nu]}}^{,\nu}).
\sectionit
$$
Taking the divergence of the left-hand-side of (3.11) and using (3.13) yields
the same as (3.14).

In the wave-zone, $T_{\mu\nu}=0$, and Eqs.(3.11) and (3.13) become
$$
(\dal+\mu^2)\psi_{\mu\nu}=0,
\sectionit
$$
$$
\psi_\mu=0.
\sectionit
$$
These equations can be obtained from the Lagrangian:
$$
{\cal L}_{\psi}={1\over 4}\psi^2_{\mu\nu,\lambda}-{1\over 2}\psi_\mu^2
-{1\over 4}\mu^2\psi_{\mu\nu}^2.
\sectionit
$$
The $\psi_{\mu\nu}$ has the spin decomposition:
$$
\psi_{\mu\nu}=1_b\oplus 1_e,
\sectionit
$$
where $1_b$ and $1_e$ denote the magnetic and electric vectors, respectively.
Only the magnetic vector propagates corresponding to a massive spin
$1^+$ pseudovector field with the propagator:
$$
\Pi={P^1_b\over k^2-\mu^2},
\sectionit
$$
where $P^1_b$ is the magnetic projection operator defined by
$$
P^1_b={1\over 2}(\theta_{\mu\rho}\theta_{\nu\sigma}-\theta_{\mu\sigma}
\theta_{\nu\rho}),
\sectionit
$$
$$
\theta_{\mu\nu}=\delta_{\mu\nu}-{k_\mu k_\nu\over k^2-\mu^2},
\sectionit
$$
and $k_\mu$ denotes the momentum four vector. The Lagrangian (3.17) is free of
ghosts, tachyons and higher-order poles$^{8}$ and the Hamiltonian is positive
and bounded from below.

The parameter $\sigma$ is fixed to be $\sigma=-1/3$ by the physical
requirement that the weak field linear approximation of the NGT field
equations is free of ghosts, tachyons and higher-order poles in the
propagator. This guarantees that the physical vacuum in the theory
is stable.

As shown by van Nieuwenhuizen$^{8}$, there exist in flat Minkowski spacetime
only two physically consistent
Lagrangians for the antisymmetric potential field $\psi_{\mu\nu}$, which
are free of ghosts, tachyons and higher-order poles in
the propagator, namely, the massless gauge invariant
theory, and the massive Proca-type theory with the Lagrangian (3.17).
Since NGT does not possess a massless gauge invariance, then the only other
possibility is that it should reduce to the physical, massive Proca-type
theory. We have, in fact, now discovered a fully geometrical NGT scheme
that fulfils this requirement.
\vskip 0.2 true in
\setsection\proclaim 4. {\bf Expansion of the Field Equations Around a Curved
Background}\par
\vskip 0.2 true in
Let us now consider the expansion of the field equations around an arbitrary
Einstein background metric. We shall introduce the notation: $g_{[\mu\nu]}
=a_{\mu\nu}$. We have
$$
g_{\mu\nu}=g_{S\mu\nu}+{}^{(1)}g_{\mu\nu}+...,\quad \Gamma^\lambda_{\mu\nu}
=\Gamma^\lambda_{S\mu\nu}+{}^{(1)}{\Gamma^\lambda_{\mu\nu}}+...,\quad
W_\mu={}^{(1)}W_\mu+{}^{(2)}W_\mu+...,
\sectionit
$$
where $g_{S\mu\nu}$ and $\Gamma^\lambda_{S\mu\nu}$ denote the Einstein
background metric tensor and connection, respectively. We also
introduce the notation for the Riemann tensor:
$$
{B^\sigma}_{\mu\nu\rho}=\Gamma^\sigma_{S\mu\nu,\rho}
-\Gamma^\sigma_{S\mu\rho,\nu}
-\Gamma^\sigma_{S\alpha\nu}\Gamma^\alpha_{S\mu\rho}
+\Gamma^\sigma_{S\alpha\rho}\Gamma^\alpha_{S\mu\nu}.
\sectionit
$$
By performing the contraction on the suffixes $\sigma$ and $\rho$,
we get the Ricci tensor, $B_{\mu\nu}={B^\alpha}_{\mu\nu\alpha}$.

Solving for the antisymmetric part of $\Gamma^\lambda_{\mu\nu}$ to first
order in $a_{\mu\nu}$ gives:
$$
^{(1)}{\Gamma^\lambda_{[\mu\nu]}}={1\over 2}g_S^{\lambda\sigma}
(\nabla_\mu a_{\sigma\nu}+\nabla_\nu a_{\mu\sigma}+\nabla_\sigma
a_{\mu\nu})+{1\over 2}\sigma({\delta^\lambda_\mu} ^{(1)}W_\nu
-{\delta^\lambda_\nu} ^{(1)}W_\mu).
\sectionit
$$
Substituting this into the field equations (2.20) ($\lambda=0$), we get to
first order in $a_{\mu\nu}$ and $W_\mu$ (we denote
$^{(1)}W_\mu$ by $W_\mu$ and ${}^{(1)}a_{\mu\nu}$ by $a_{\mu\nu}$):
$$
B_{\mu\nu}=0,
\sectionit
$$
$$
\nabla^\sigma a_{\mu\sigma}={3\over 2}\sigma W_\mu,
\sectionit
$$
$$
(\dal_S+\mu^2)a_{\mu\nu}-2g_S^{\lambda\sigma}g_S^{\alpha\beta}
B_{\alpha\nu\lambda\mu}a_{\sigma\beta}
=\kappa\nabla_{[\mu}W_{\nu]},
\sectionit
$$
where, as in the previous section, $\kappa=\sigma+{4\over 3}$. Moreover,
we have $\dal_S=\nabla^\sigma\nabla_\sigma,
\nabla^\sigma=g_S^{\sigma\alpha}\nabla_\alpha$ and:
$$
g_S^{\lambda\alpha}g_{S\beta\lambda}=\delta^\alpha_\beta.
\sectionit
$$
We also employed the result:
$$
\nabla^\sigma\nabla_\mu a_{\sigma\nu}
=\nabla_\mu\nabla^\sigma a_{\sigma\nu}-g_S^{\lambda\sigma}g_S^{\alpha\beta}
(B_{\alpha\sigma\lambda\mu}a_{\beta\nu}
+B_{\alpha\nu\lambda\mu}a_{\sigma\beta})
$$
$$
=\nabla_\mu\nabla^\sigma a_{\sigma\nu}-g_S^{\lambda\sigma}g_S^{\alpha\beta}
B_{\alpha\nu\lambda\mu}a_{\sigma\beta},
\sectionit
$$
where in the last line we used Eq.(4.4).

Taking the divergence of (4.6), we get
$$
\tau\dal_S W_\mu-2\nabla^\nu(g_S^{\lambda\sigma}g_S^{\alpha\beta}
B_{\alpha\nu\lambda\mu}a_{\sigma\beta})
+{3\over 2}\sigma\mu^2 W_\mu+\nabla^\nu(Ba)_{\mu\nu}=0,
\sectionit
$$
where
$$
\nabla^\nu\dal_S a_{\mu\nu}=\dal_S\nabla^\nu a_{\mu\nu}+\nabla^\nu
(Ba)_{\mu\nu},
\sectionit
$$
and $\tau=2(\sigma+{1\over 3})$. Also, $(Ba)_{\mu\nu}$ denotes additional terms
involving products of the Riemann tensor and $a_{\mu\nu}$.
We choose $\tau=0$ to give
$$
W_\mu=-{1\over \mu^2}
\biggl[\nabla^\nu\biggl(4g_S^{\lambda\sigma}g_S^{\alpha\beta}
B_{\alpha\nu\lambda\mu}a_{\sigma\beta}
-2(Ba)_{\mu\nu}\biggl)\biggr].
\sectionit
$$

{}From (4.5) and (4.11), we get the field equations:
$$
\nabla^\sigma a_{\mu\sigma}={1\over 2\mu^2}
\biggl[\nabla^\nu\biggl(4g_S^{\lambda\sigma}g_S^{\alpha\beta}
B_{\alpha\nu\lambda\mu}a_{\sigma\beta}
-2(Ba)_{\mu\nu}\biggr)\biggr],
\sectionit
$$
$$
(\dal_S+\mu^2)a_{\mu\nu}=M_{\mu\nu},
\sectionit
$$
where
$$
M_{\mu\nu}=2g_S^{\lambda\sigma}g_S^{\alpha\beta}
B_{\alpha\nu\lambda\mu}a_{\sigma\beta}
+{1\over \mu^2}
\nabla_{[\nu}\nabla^\rho\biggl[4g_S^{\lambda\sigma}g_S^{\alpha\beta}
B_{\alpha\rho\lambda\mu]}a_{\sigma\beta}-2(Ba)_{\mu]\rho}\biggr].
\sectionit
$$

As before, $W_\mu$ does not propagate and there is no coupling to unphysical
modes through the effective source tensor formed from the Riemann tensor and
$a_{\mu\nu}$.

The energy associated with the flux of gravitational waves, calculated in the
wave-zone for $r\rightarrow \infty$, is positive definite.
\vskip 0.2 true in
\setsection\proclaim 5.
{\bf Time-Dependent Spherically Symmetric Field Equations}
\par\vskip 0.2 true in
In the case of a spherically symmetric field,
Papapetrou has derived the canonical form of $g_{\mu\nu}$ in NGT$^{9}$:
$$
g_{\mu\nu}=\left(\matrix{-\alpha&0&0&w\cr
0&-\beta&f\hbox{sin}\theta&0\cr 0&-f\hbox{sin}\theta&
-\beta\hbox{sin}^2
\theta&0\cr-w&0&0&\gamma\cr}\right),
\sectionit
$$
where $\alpha,\beta,\gamma$ and $w$ are functions of $r$ and $t$. The
tensor $g^{\mu\nu}$ has the components:
$$
g^{\mu\nu}=\left(\matrix{{\gamma\over w^2-
\alpha\gamma}&0&0&{w\over w^2-\alpha\gamma}\cr
0&-{\beta\over \beta^2+f^2}&{f\hbox{csc}\theta\over \beta^2+f^2}&0\cr
0&-{f\hbox{csc}\theta\over \beta^2+f^2}&-{\beta\hbox{csc}^2\theta\over
\beta^2+f^2}&0\cr-{w\over w^2-\alpha\gamma}&0&0&-{\alpha\over
w^2-\alpha\gamma}\cr}\right).
\sectionit
$$
For the theory in which there is no NGT magnetic monopole charge, we have
$w=0$ and only the $g_{[23]}$ component of $g_{[\mu\nu]}$ survives.

The line element for a spherically symmetric body is given by
$$
ds^2=\gamma(r,t)dt^2-\alpha(r,t)dr^2-\beta(r,t)(d\theta^2+\hbox{sin}^2\theta
d\phi^2).
\sectionit
$$
We have
$$
\sqrt{-g}=\hbox{sin}\theta[\alpha\gamma(\beta^2+f^2)]^{1/2}.
\sectionit
$$

The vector $W_\mu$ can be determined from Eq.(2.12):
$$
W_\mu=-2k_{\mu\rho}{{\bf g}^{[\rho\sigma]}}_{,\sigma},
\sectionit
$$
where $k_{\mu\nu}$ is defined by $k_{\mu\alpha}g^{(\mu\beta)}
=\delta_\alpha^\beta$. For the static spherically symmetric field with
$w=0$ it follows from (5.5) that $W_\mu=0$.

The field equations (2.20) in empty space are given by
$$
R_{11}(\Gamma)=-{1\over 2}A^{''}-{1\over 8}[(A^\prime)^2+4B^2]
+{\alpha^\prime A^\prime\over 4\alpha}+{\gamma^\prime\over
2\gamma}\biggl({\alpha^\prime\over 2\alpha}
-{\gamma^\prime\over 2\gamma}\biggr)
$$
$$
-\biggl({\gamma^\prime\over 2\gamma}\biggr)^\prime
+{\partial\over \partial t}\biggl({{\dot \alpha}\over 2\gamma}\biggr)
+{{\dot \alpha}\over 2\gamma}\biggl({{\dot\gamma}\over 2\gamma}
-{{\dot\alpha}\over 2\alpha}
+{1\over 2}{\dot A}\biggr)=-\lambda \alpha+{1\over 4}\mu^2{\alpha f^2\over
\beta^2+f^2},
\sectionit
$$
$$
{1\over \beta}R_{22}(\Gamma)={1\over
\beta}R_{33}(\Gamma)\hbox{cosec}^2\theta={1\over\beta}+
{1\over \beta}\biggl({2fB-\beta A^\prime\over 4\alpha}\biggr)^\prime
+{2fB-\beta A^\prime\over 8\alpha^2\beta\gamma}(\alpha^\prime\gamma
+\gamma^\prime\alpha)
$$
$$
+{B(fA^\prime+2\beta B)\over 4\alpha\beta}
-{1\over \beta}{\partial\over \partial t}\biggl({2fD-\beta{\dot A}\over
4\gamma}\biggr)
-{2fD-\beta{\dot A}\over 8\alpha\gamma^2\beta}({\dot\alpha}\gamma
+\gamma{\dot \alpha})
$$
$$
-{D\over 4\gamma\beta}(f{\dot A}+2\beta D)=-\lambda-{1\over 4}\mu^2{f^2\over
\beta^2+f^2},
\sectionit
$$
$$
R_{00}(\Gamma)
=-{1\over 2}{\ddot A}-{1\over 8}({\dot A}^2+4D^2)+{{\dot\gamma}\over
4\gamma}{\dot A}+{{\dot\alpha}\over 2\alpha}\biggl({{\dot\gamma}\over 2\gamma}
-{{\dot\alpha}\over 2\alpha}\biggr)
$$
$$
-{\partial\over \partial t}\biggl({{\dot\alpha}\over 2\alpha}\biggr)
+\biggl({\gamma^\prime\over 2\alpha}\biggr)^\prime
+{\gamma^\prime\over 2\alpha}\biggl({\alpha^\prime\over 2\alpha}
-{\gamma^\prime\over 2\gamma}+{1\over 2}A^\prime\biggr)=\lambda\gamma
-{1\over 4}\mu^2{\gamma f^2\over \beta^2+f^2},
\sectionit
$$
$$
R_{[10]}(\Gamma)=0,
\sectionit
$$
$$
R_{(10)}(\Gamma)=-{1\over 2}{\dot A}^\prime+{1\over 4}A^\prime
\biggl({{\dot\alpha}\over \alpha}-{1\over 2}{\dot A}\biggr)+{1\over 4}
{\gamma^\prime{\dot A}\over \gamma}-{1\over 2}BD=0,
\sectionit
$$
$$
R_{[23]}(\Gamma)=\hbox{sin}\theta\biggl[\biggl({fA^\prime+2\beta B\over
4\alpha}\biggr)^\prime+{1\over 8\alpha}(fA^\prime+2\beta B)
\biggl({\alpha^\prime\over \alpha}+{\gamma^\prime\over\gamma}\biggr)
$$
$$
-{B\over 4\alpha}(2fB-\beta A^\prime)-{1\over 8\gamma}(f{\dot A}+2\beta D)
\biggl({{\dot\gamma}\over \gamma}+{{\dot\alpha}\over \alpha}\biggr)
$$
$$
-{\partial\over \partial t}\biggl({f{\dot A}+2\beta D\over 4\gamma}\biggr)
+{D\over 4\gamma}(2fD-\beta{\dot A})\biggr]=\biggl[\lambda f
-{1\over 4}\mu^2f\biggl(1+{\beta^2\over \beta^2+f^2}\biggr)\biggr]
\hbox{sin}\theta.
\sectionit
$$
Here, prime denotes differentiation with respect to $r$, ${\dot A}=\partial A/
\partial t$, and we have used the notation:
$$
A=\hbox{ln}(\beta^2+f^2),
\sectionit
$$
$$
B={f\beta^\prime-\beta f^\prime\over
\beta^2+f^2},
\sectionit
$$
$$
D={{\dot\beta}f-{\dot f}\beta\over \beta^2+f^2}.
\sectionit
$$

We can rearrange (5.10) to give
$$
{\dot\alpha}={\alpha\over A^\prime}\biggl[{1\over 2}A^\prime {\dot A}
+2{\dot A}^\prime-{\gamma^\prime\over \gamma}{\dot A}+2BD\biggr].
\sectionit
$$
Let us choose $\beta=r^2$ and substitute this into (5.15):
$$
{\dot \alpha}=\alpha\biggl({r^4+f^2\over 2r^3+ff^\prime}\biggr)
\biggl[{f^\prime{\dot f}(r^4-3f^2)\over (r^4+f^2)^2}-{8f{\dot f}r^3\over
(r^4+f^2)^2}+{2(f^\prime{\dot f}+f{\dot f}^\prime)\over r^4+f^2}
$$
$$
-\biggl({\gamma^\prime\over\gamma}\biggr)
\biggl({f{\dot f}\over r^4+f^2}\biggr)\biggr].
\sectionit
$$
When we choose $f=0$, we obtain Birkhoff's theorem in EGT i.e., the
time-dependent spherically symmetric solution must be stationary
and it is the static Schwarzschild solution for $r > 2m\,^{10}$.

To see whether there exists a Birkhoff theorem for the NGT empty space field
equations, we shall consider the asymptotic large $r$, static approximation of
the field equation (5.11) with $\lambda=0, \beta=r^2$ and
$\alpha\sim \gamma\sim 1$:
$$
f''-{2\over r}f'-\mu^2f=0.
\sectionit
$$
This has the solution valid for $\mu m << 1$:
$$
f=C\hbox{exp}(-\mu r)(1+\mu r),
\sectionit
$$
where $C$ is a constant. To this order of approximation the invariant:
$F=g_{[23]}g^{[23]}$ is
given by
$$
F\sim\hbox{const.}\,\hbox{exp}(-\mu r)\biggl({\mu\over r}+{1\over r^2}\biggr).
\sectionit
$$

For $r\rightarrow\infty$, we see from (5.15) that ${\dot\alpha}$ vanishes
rapidly and no monopole radiation will be observed in the wave-zone for large
$r$. Asymptotically the solution is the static solution of the field equations,
yielding an extended Birkhoff theorem.
\vskip 0.2 true in
\setsection\proclaim 6.
{\bf Static Spherically Symmetric Solution}\par
\vskip 0.2 true in
Let us assume the short-range approximation for which the $\mu^2$ contributions
in the vacuum field equations can be neglected and we assume that
$\mu^{-1} > 2m$. We then obtain the static, spherically symmetric Wyman
solution$^{11}$:
$$
\gamma=\hbox{exp}(\nu),
\sectionit
$$
$$
\alpha=m^2(\nu^\prime)^2\hbox{exp}(-\nu)(1+s^2)
(\hbox{cosh}(a\nu)-\hbox{cos}(b\nu))^{-2},
\sectionit
$$
$$
f=[2m^2\hbox{exp}(-\nu)(\hbox{sinh}(a\nu)\hbox{sin}(b\nu)+s(1-\hbox{cosh}(a\nu)
\hbox{cos}(b\nu))](\hbox{cosh}(a\nu)-\hbox{cos}(b\nu))^{-2},
\sectionit
$$
where $\nu$ is implicitly determined by the equation:
$$
\hbox{exp}(\nu)(\hbox{cosh}(a\nu)-\hbox{cos}(b\nu))^2{r^2\over 2m^2}=
\hbox{cosh}(a\nu)\hbox{cos}(b\nu)-1+s\hbox{sinh}(a\nu)\hbox{sin}(b\nu),
\sectionit
$$
and $s$ is a dimensionless constant of integration.

We find for $2m/r << 1$ and $0 < sm^2/r^2 < 1$ that the metric
takes the Schwarzschild form:
$$
\gamma=\alpha^{-1}=1-{2m\over r}.
\sectionit
$$

Near $r=0$ we can develop expansions where $r/m < 1$ and $0 < \vert s\vert
<1$. The leading terms are$^{7}$:
$$
\gamma=\gamma_0+{\gamma_0(1+{\cal O}(s^2))\over 2\vert s\vert}\biggl({r\over
m}\biggr)^2 + {\cal O}\biggl(\biggl({r\over m}\biggr)^4\biggr),
\sectionit
$$
$$
\alpha={4\gamma_0(1+{\cal O}(s^2))\over s^2}\biggl({r\over m}\biggr)^2
+{\cal O}\biggl(\biggl({r\over m}\biggr)^4\biggr),
\sectionit
$$
$$
f=m^2\biggl(4-{\vert s\vert\pi\over 2}+s\vert s\vert+{\cal O}(s^3)\biggr)
+{\vert s\vert+s^2\pi/8+{\cal O}(s^3)\over 4}r^2+{\cal O}(r^4),
\sectionit
$$
$$
\gamma_0=\hbox{exp}\biggl(-{\pi+2s\over \vert s\vert}+{\cal O}(s)\biggr)...\,.
\sectionit
$$
These solutions clearly illustrate
the non-analytic nature of the limit $s\rightarrow 0$ in the strong
gravitational field regime for $r < 2m$.

The singularity caused by the vanishing of $\alpha(r)$ at
$r=0$ is a {\it coordinate} singularity, which can be removed by
transforming to another coordinate frame of reference. The curvature
invariants do not, of course, contain any coordinate singularities.
We can transform to a coordinate frame in which the spacetime
$(M,g_{\mu\nu})$ is completely free of singularities.

The NGT curvature invariants such as the generalized Kretschmann scalar:
$$
K=R^{\lambda\mu\nu\rho}R_{\lambda\mu\nu\rho}
\sectionit
$$
are finite$^{7}$.

The static spherically symmetric solution is everywhere non-singular and
there is no event horizon at $r=2m$. A black hole is replaced in the theory
by a superdense object which can be stable for an arbitrarily large
mass$^{7,12}$.

A solution of the vacuum field equations remains to be found
when the $\mu^2$ contributions are kept in (5.6)-(5.11). It is
expected that this solution will retain the non-singular properties of the
short-range approximation solution, for the $\mu^2$ contributions will
damp out the effects of $g_{[\mu\nu]}$ at large distances.
\vskip 0.2 true in
\setsection\proclaim 7.
{\bf Conclusions}\par
\vskip 0.2 true in
We have succeeded in deriving a geometrical version of NGT, based on the
Lagrangian density:
$$
{\cal L}=\sqrt{-g}[R(W)-2\lambda-{1\over 4}\mu^2g^{\mu\nu}g_{[\nu\mu]}
-{1\over 6}g^{\mu\nu}W_\mu W_\nu],
\sectionit
$$
which yields a stable, ghost and tachyon free linear approximation, when
the field equations are expanded about Minkowski spacetime or about an
arbitrary Einstein background metric.

In the short-range approximation, in which $\mu^{-1}$ is large compared to
the length scales being investigated, a static spherically symmetric
solution can be obtained from the field equations which is regular everywhere
in spacetime and which does not contain an event horizon at $r=2m$. The
black hole that is predicted by the collapse of a sufficiently massive
star in EGT is replaced by a superdense object that is stable for an
arbitrarily large mass.

Attempts must be made to investigate solutions of the field equations
valid for any value of the range parameter $\mu^{-1}$, so that the regularity
of the solutions can be established, in general. However, since the
$g_{[\mu\nu]}$ solutions are damped out for $r\rightarrow \infty$, we do not
expect that the regularity properties of the Wyman solution to be modified.

An important result of this study is that a classical
theory of gravity can be formulated, which has a physically consistent
perturbative expansion for weak fields, and does not have singularities and
black holes. It satisfies the standard gravitational experimental tests.

Recently, a solution has been published for inhomogeneous gravitational
collapse of a matter cloud with a general form of matter, which leads
to a naked singularity$^{13}$. The collapse is related to the choice of
initial data for the Einstein field equations, and the naked singularity
would occur in generic situations based on regular initial conditions
satisfying the weak energy condition. This result would lead to the
demise of EGT, for it represents a local violation of the Cauchy data for
collapse; it would provide a strong motivation for seriously considering
a classical gravity theory such as NGT with everywhere regular solutions
of the field equations.

Because there is no event horizon in the static spherically symmetric solution,
we can resolve the problem of information loss$^{14}$ associated with black
holes at a classical level.
\vskip 0.2 true in
{\bf Acknowledgements}
\vskip 0.2 true in
I thank M. Clayton, N. J. Cornish, J. L\'egar\'e and P. Savaria for helpful
and stimulating discussions. I thank the Natural Sciences and Engineering
Research Council of Canada for the support of this work.
\par\vfil\eject
\centerline{\bf References}
\vskip 0.2 true in
\item{1.}{J. W. Moffat, Phys. Rev. D {\bf 19}, 3554 (1979); D {\bf 19},
3562 (1979); J. W. Moffat, J. Math. Phys. {\bf 21}, 1978 (1980); R. B.
Mann and J. W. Moffat, Phys. Rev. D {\bf 26}, 1858 (1982);
J. W. Moffat, Found. Phys. {\bf 14}, 1217 (1984); J. W. Moffat, Phys.
Rev. D {\bf 35}, 3733 (1987). For a review of NGT, see: J. W. Moffat,
Proceedings of the
Banff Summer Institute on Gravitation, Banff, Alberta, August 12-25, 1990,
edited by R. Mann and P. Wesson, World Scientific, p.523, 1991.}
\item{2}{A. Einstein, The Meaning of Relativity, Menthuen and Co. London,
U.K. fifth edition, Appendix II, 1951. Einstein used the nonsymmetric field
structure to derive a unified field theory of gravitation and
electromagnetism. The nonsymmetric field equations cannot produce
a consistent physical theory of electromagnetism. We use the nonsymmetric
field structure to descibe a non-Riemannian theory of gravitation.}
\item{3.}{R. B. Mann and J. W. Moffat, J. Phys. A {\bf 14}, 2367 (1981);
{\it ibid}, {\bf 15}, 1055(E) (1982).}
\item{4.}{T. Damour, S. Deser and J. McCarthy, Phys. Rev. D {\bf 47}, 1541
(1993).}
\item{5.}{J. W. Moffat, University of Toronto preprint, UTPT-94-28, gr-qc/
9411006.}
\item{6.}{K. S. Stelle, Phys. Rev. D {\bf 16}, 953 (1977); E. Sezgin and P. van
Nieuwenhuizen, Phys. Rev. D {\bf 21}, 3269 (1980).}
\item{7.}{N. J. Cornish and J. W. Moffat, Phys. Letts. {\bf 336B}, 337 (1994);
University of Toronto preprint, UTPT-94-08, 1994, gr-qc/9406007. To be
published
in J. Math. Phys.; I. Yu.Sokolov, University of Toronto preprint, UTPT-94-23,
1994.}
\item{8.}{P. van Nieuwenhuizen, Nucl. Phys. B {\bf 60}, 478 (1973).}
\item{9.}{A. Papapetrou, Proc. Roy. Ir. Acad. Sec. A {\bf 52}, 69 (1948).}
\item{10.}{G. D. Birkhoff, Relativity and Modern Physics, Harvard
University Press, Cambridge, Massachusetts, 1923.}
\item{11.}{M. Wyman, Can. J. Math. {\bf 2}, 427 (1950).}
\item{12.}{N. J. Cornish, University of Toronto preprint, UTPT-94-10, 1994,
gr-qc/9405065; revised November, 1994.}
\item{13.}{P. S. Joshi and I. H. Dwivedi, Commun. Math. Phys. {\bf 146},
333 (1992); to appear in Commun. Math. Phys. (1994); P. S. Joshi, Global
Aspects in Gravitation and Cosmology, Clarendon Press, Oxford, 1994.}
\item{14.}{S. Hawking, Phys. Rev. D {\bf 14}, 2460 (1976); Commun. Math.
Phys. {\bf 87}, 395 (1982).}

\end